\documentclass[10pt,twocolumn]{article}

\usepackage{graphics}
\usepackage{overcite}
\usepackage{times}

\begin{document}

\def\etal{{\it{}et~al.}}
\def\eref#1{(\protect\ref{#1})}
\def\d{{\rm d}}
\def\columnfigure#1{\resizebox{\linewidth}{!}{\includegraphics{#1}}}

\title{Simple models of evolution and extinction}
\author{M. E. J. Newman\\
{\normalsize\it Santa Fe Institute, 1399 Hyde Park Road, Santa Fe, NM 87501}}
\date{}
\maketitle

\begin{abstract}
  This article gives a brief introduction to the mathematical modeling of
  large-scale biological evolution and extinction.  We give three examples
  of simple models in this field: the coevolutionary avalanche model of Bak
  and Sneppen, the environmental stress model of Newman, and the increasing
  fitness model of Sibani, Schmidt, and Alstr\o{}m.  We describe the
  features of real evolution which these models are intended to explain and
  compare the results of simulations against data drawn from the fossil
  record.
\end{abstract}

\section{Introduction}
Throughout the 3 billion year history of life on the Earth the processes of
evolution and extinction have been inextricably linked.  Species survive on
average only about 10 million years before they become extinct, so that
almost every species that has ever lived is extinct today.  This high
turnover of species has played a crucial role in long-term evolution
because it is the removal of one species which makes way for the evolution
of another.  The classic example is that of the dinosaurs, whose extinction
at the end of the Cretaceous period 65~million years ago cleared the way
for the subsequent dominance of the mammals and eventually the evolution of
the human race.

Most of our knowledge about prehistoric life comes from the fossil record.
Traditionally, fossil studies have focused on the evolution of individual
species or groups of species, or on prominent prehistoric events such as
mass extinctions or adaptive radiations (the evolution and spread of
species to occupy new niches in the ecosystem).  However, in the last ten
years or so, with the availability of extensive computer databases of
fossil species, researchers have also started to look at large-scale
patterns in the fossil record, such as the distribution of the sizes of
extinction events, and the distribution of the lifetimes of species or
groups of species.  These studies have led to the suggestion of a variety of
new mechanisms which may affect evolution and extinction on long time
scales, and of mathematical models incorporating these mechanisms which can
mimic some aspects of the development of life.  In the following we describe
some of the patterns seen in the fossil data and some of the models which
have been proposed to explain them.

\section{The fossil data}
Currently available databases of the fossil record represent about a
quarter of a million species, mostly marine animals, which are usually
grouped either into genera or into families (the two levels of the Linnean
hierarchy immediately above species).  The reason for this grouping is that
there are not enough fossils of most individual species to make meaningful
estimates of when they first appeared and when they became extinct.  By
grouping them into genera and families we increase the number of fossils
per group and thereby the accuracy of origination and extinction dates.

Dating is usually done to the nearest stratigraphic stage.  Stages are
irregular intervals of time of average duration of about seven million
years, which are based on easily identifiable geological features.  Almost
all the available fossil data come from the Phanerozoic eon, the last 540
million years, during which multicellular life has dominated the planet.
There are 77 stages in the Phanerozoic.

One of the most striking proposals that has been put forward in the last
few years is that some distributions of fossil quantities may follow power
laws in which the probability $p(x)$ of measuring a value $x$ for a
particular quantity satisfies
\begin{equation}
p(x) \sim x^{-\alpha}.
\end{equation}
When such a distribution is plotted on logarithmic scales, one obtains a
straight line
\begin{equation}
\log p(x) \sim -\alpha\log x,
\end{equation}
with slope $-\alpha$.

\begin{figure}
\columnfigure{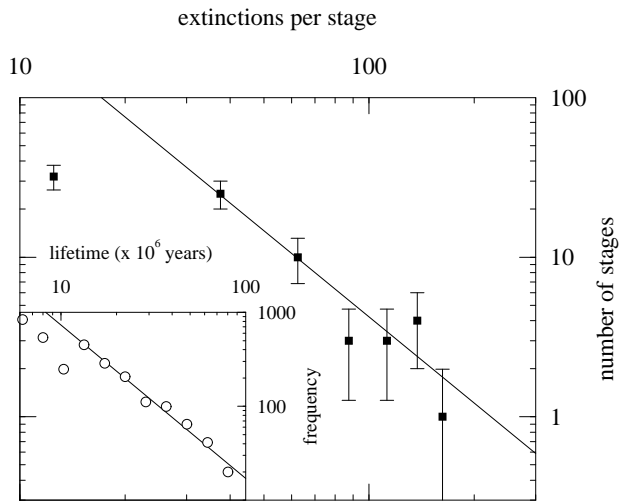}
\caption{Histogram of the number of families of fossil marine
  species becoming extinct per stratigraphic stage of the Phanerozoic.
  Inset: histogram of the lifetimes of genera in the fossil record.  The
  data are taken from the compilation by
  Sepkoski.\protect\cite{Sepkoski93}}
\label{extinction}
\end{figure}

Figure~\ref{extinction} shows a histogram of the number of families
becoming extinct per stratigraphic stage on a log-log scale.  The horizontal
axis measures the number $x$ of families that became extinct in any given
stage of the Phanerozoic, and the vertical axis measures the number of
stages in which $x$ took that value.  The histogram is clearly skewed
heavily to the right---there are many stages in which a few families became
extinct, and few in which many became extinct.  It has been
suggested\cite{Kauffman93,SB96,Newman96} that this histogram follows a
power law with a slope of about $-2$.  In the inset of
Figure~\ref{extinction} we show a histogram of the lifetimes of genera,
which also appears to follow a power law,\cite{SBFJ95,NS99} with a slope in
this case of about $-\frac32$.  In the following sections we look at some
simple models which have been proposed as possible explanations for the
generation of power laws such as these.

\section{The model of Bak and Sneppen}
The model which has probably generated the most excitement in this field,
and which must be credited with stimulating a large part of the recent
interest in evolution modeling within the computer simulation community, is
the self-organized critical evolution model of Bak and Sneppen.\cite{BS93}
The basic idea behind this model is that of the {\it fitness landscape}.

It was the influential British biologist Sewell Wright who first proposed
that evolution be viewed as a combinatoric optimization process on a rugged
landscape,\cite{Wright67} similar to the satisfiability problems of
computer science\cite{Papadimitriou94} or spin glasses in
physics.\cite{FH91} Organisms or species can be thought of as having a
scalar ``fitness,'' usually denoted by $W$, which measures their
reproductive success.  Species with higher reproductive success have more
offspring in the next generation and dominate over species with lower
reproductive success.  For every possible genotype of an organism, that is,
for every possible sequence of its DNA, there is an associated value of $W$
which is the fitness of the organism that has that gene sequence.  The
mapping from genotype to fitness is the fitness landscape.  The landscape
exists in a very high dimensional space similar to the state space of a
physical system such as a spin system.

Evolution serves to move species on the fitness landscape.  Because species
with higher fitness are favored over those with lower fitness, a mutant
strain of organism which finds itself at a higher point on the fitness
landscape will dominate over its ancestral strain and over time the
population will shift to the fitter genotype.  Thus, under the influence of
repeated mutation and selection, species tend to move ``uphill'' on the
fitness landscape, only stopping when they reach a local maximum or peak on
the landscape.  The peaks on a landscape represent all the possible stable
species.  (Ideas inspired by this view of evolution have been used to
formulate new optimization methods in computer science.  These methods
typically go by the name of {\it genetic algorithms}\cite{Mitchell96} or
{\it genetic programming}.\cite{Koza92})

Life would be boring in an ecosystem in which all species simply walked
uphill on their own fitness landscape until they reached a local peak.
Once everyone found their peak, evolution would stop.  This situation is
called {\it Nash equilibrium}.  There are a variety of reasons why this
situation does not happen in real evolution.  First, there may be
perturbations from the environment, pressures such as changing climate or
changing food supply, which affect the shape of the fitness landscape and
force species which were previously stable to evolve into new forms.  Even
in the absence of such perturbations however, evolution may still
occur.\cite{KJ91} It is possible for a stable population to evolve if one
of the members of that population undergoes a large mutation, or a rapid
sequence of smaller ones, which moves it so far on the fitness landscape
that it finds itself in the basin of attraction of a new fitness peak.
Another possibility is that evolution takes place because of interactions
between species.  Species are not independent; the evolution of one can
affect the fitness of another.  For example, if you prey on a certain
animal which evolves to fly in order to escape you, then you had better
evolve to fly too, or learn to eat something else, or you are likely to die
out.  Thus, the evolution of one species affects the shape of the fitness
landscape of the others with which it interacts.  This process is called
{\it coevolution}.

\begin{figure}
\columnfigure{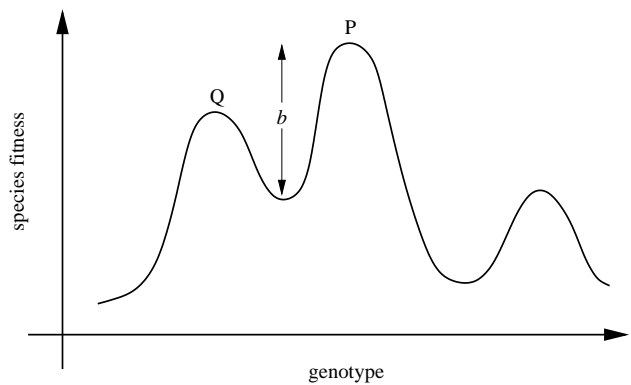}
\caption{A schematic representation of the ``fitness barrier''~$b$ which
  a species at peak $P$ must surmount in order to evolve to a new fitness
  peak $Q$.  Note that the new peak need not be higher than the old one.}
\label{barrier}
\end{figure}

Bak and Sneppen\cite{BS93} incorporated these ideas into a simple model of
evolution as follows.  Suppose we have a certain number $N$ of species in
an ecosystem, each of which is localized around a peak on its own fitness
landscape.  Each species interacts with a number of others, which can be
chosen in a variety of ways.  The simplest way is to place the species on a
lattice and have each interact with its nearest neighbors.  Nothing will
happen to any of the species as long as they remain at their local peaks.
However, every once in a while, a large mutation or sequence of mutations
will take a species from its local peak over to the basin of attraction of
another peak, and so cause it to evolve.  Bak and Sneppen represented the
ease with which this ``excitation'' could take place by a ``fitness
barrier'' $b_i$ for the mutation of the $i$th species (see
Figure~\ref{barrier}), analogous to the energy barrier which a physical
system has to cross to move from one local energy minimum to another on a
rugged energy landscape.  The species which has the lowest barrier to
mutation is assumed to be the one that evolves first.

Here is where coevolution comes in.  When a species evolves by crossing its
fitness barrier, it affects the shape of the fitness landscapes of those
species with which it interacts.  Bak and Sneppen made the simplifying
assumption that the fitness landscape of the neighboring species is
completely randomized.  These neighboring species, which were previously at
a comfortable local peak, now find themselves (most probably) not at a peak
at all, and so evolve again until they reach a new peak, with a new fitness
barrier.  This process is represented in the Bak--Sneppen model by choosing
a new value at random for the fitness barrier of each neighboring species.
But the process stops here: it is assumed that the neighbors of the
neighbors do not also evolve.  The next species to evolve are the one with
the next lowest fitness barrier and its neighbors.  Thus the entire model
can be summarized as follows:
\begin{enumerate}
\item $N$ species are placed on a lattice and each is given a fitness
  barrier $b_i$, which initially is chosen at random.  Bak and Sneppen used
  uniform random numbers between zero and one, and this choice seems as
  good as any.
\item The species with the lowest barrier is found, and its fitness barrier
  is replaced by a new value, again chosen at random.
\item The nearest neighbors of this species on the lattice are given new
  random barrier values also.
\item Repeat from step~2.
\end{enumerate}

And that is the entire model.  So what does the model do?  Well, initially,
the dynamics tends to remove all the low-lying barriers from the system and
replace them with higher ones, producing a ``gap'' at the bottom end of the
barrier distribution---a range from zero up to some finite value in which
none of the barriers fall.  However, as time goes by, the gap becomes
larger and the probability that a new randomly chosen barrier value falls
in this gap increases proportionately.  Depending on the coordination
number of the lattice (the number of nearest neighbors of each site), the
system will reach a critical point where each time one species is removed
from the gap we put another one in, and the system reaches a dynamic
equilibrium in which the gap no longer grows.

Bak and Sneppen observed the lengths of the sequences of moves from the
moment when a species appears in the gap until the last one is removed.
(It is usually fairly clear from the distribution of barrier values where
the edge of the gap is---the distribution drops off very sharply there.  As
Bak and Sneppen showed however, you can obtain good results even if you
only get the position of the edge approximately correct.)  These sequences
they called {\it coevolutionary avalanches}, a name adopted from the
writings of Kauffman.\cite{Kauffman93} These avalanches are, in a sense,
all the result of one initial evolutionary event in which a species
spontaneously mutates to a new genotype which has a barrier value which
falls in the gap at the bottom of the distribution.  As the gap becomes
larger, the lengths of the avalanches increase, until, at the critical
point, the average avalanche length diverges, resulting in a scale-free
(that is, power-law) distribution of avalanche sizes.  Bak and Sneppen
speculated that a power-law distribution of coevolutionary avalanches could
be the cause of a power-law distribution of extinction sizes in the fossil
record: when many species evolve to new forms, all the ancestral forms die
out, causing a mass extinction.

\begin{figure}
\columnfigure{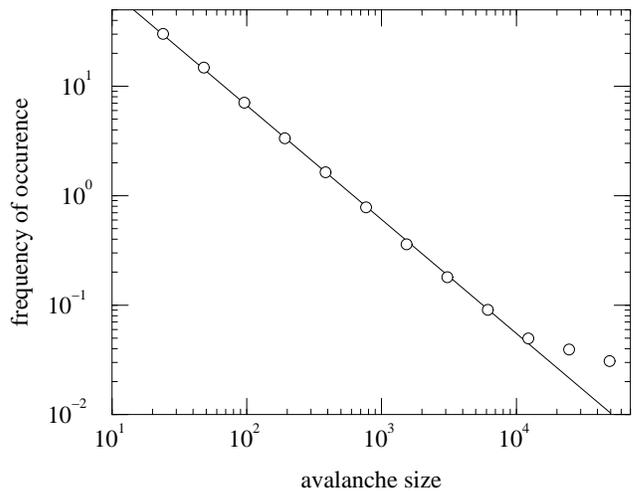}
\caption{Histogram of the sizes of coevolutionary avalanches in a
  simulation of the Bak--Sneppen model\protect\cite{BS93} with $N=100$
  species on a one-dimensional lattice.  The distribution is close to
  power-law in form and for this simulation has a measured exponent of
  $-1.04\pm0.01$ (the solid line).}
\label{bsaval}
\end{figure}

In Figure~\ref{bsaval} we show a histogram of the sizes of avalanches in a
simulation of the Bak--Sneppen model after it has come to equilibrium and
indeed we do see that the distribution has the form of a power law.  In
this case the simulation was performed on a one-dimensional lattice and the
exponent of the power-law is about $-1$.  This exponent varies with the
dimensionality of the lattice, but never gets steeper than $-\frac32$,
which is still some way from the value of $-2$ estimated from the fossil
data.\cite{Newman96} This difference is one of the main drawbacks of the
Bak--Sneppen model.  Another is that the mechanism it proposes whereby
ancestral species are wiped out {\it en masse\/} by their large-scale
evolution into new forms is not thought by paleontologists to be a
realistic view of what happens in nature.  In fact, most mass extinction
events are believed to be caused by stresses on the ecosystem coming from
external causes, such as drops in sea level,\cite{Hallam89} impacts of
extraterrestrial bodies,\cite{Alvarez87} climate change,\cite{Stanley88} or
changes in the level of oxygen in the oceans.\cite{WB84} Both this issue,
and the issue of the value of the exponent have been addressed by another
simple model of extinction proposed by Newman (that's me).

\section{Newman's extinction model}
Newman\cite{Newman97} has proposed a model of extinction that takes an
approach diametrically opposite to that of Bak and Sneppen.  Where the
Bak--Sneppen model assumes that extinction is caused entirely by
(co)evolutionary effects, Newman's model assumes that it is caused entirely
by stresses on the ecosystem from external sources.  In fact, there is no
interaction between species at all in this model.  The reason why large
numbers of species become extinct simultaneously is not because they
interact with one another, but because they all feel the same stresses at
the same time.

The model works like this.  We again assume a fixed number $N$ of species,
each characterized by a single scalar $x_i$ which is the threshold amount
of stress that the species can withstand before it becomes extinct.  Stress
is represented by a noise variable $\eta(t)$, which fluctuates randomly
with time $t$.  The source of the stresses is not specified in the
model---only the magnitude of the stress matters.  The dynamics of the
model is simple: if at any time the stress $\eta(t)$ is numerically greater
than the threshold $x_i$ that species $i$ can withstand, then this species
becomes extinct at time $t$.  The niches vacated by extinct species are
repopulated by new ones which have randomly chosen thresholds $x_i$.  The
distribution of the values of stress $\eta$ is usually chosen to be some
decreasing function of $\eta$, so that large stresses are less common than
small ones.

In fact, this is not quite all there is to Newman's model.  If it were,
then the dynamics of the model would stagnate quickly once all the species
with low thresholds were removed, leaving only those species with
thresholds sufficiently high that they cannot easily be reached by stresses
of typical size.  To prevent this happening, Newman also included an
evolution mechanism in the model, whereby species are occasionally chosen
at random and their thresholds changed to new randomly chosen values.  This
mechanism means that there is always an influx of new species with low
thresholds to feed the extinction process.

The model can be summarized as follows:
\begin{enumerate}
\item Each of the $N$ species is given a threshold value which is initially
  chosen at random, usually from a uniform distribution between zero and
  one.
\item A random number $\eta$ is chosen from some distribution $p_{\rm
    stress}(\eta)$ to represent the current stress level.  All species $i$
  for which $x_i<\eta$ are wiped out and are replaced by new species with
  randomly chosen thresholds $x_i$ (which may be less than $\eta$).
\item A small fraction $f$ of the species are picked at random and
  ``evolved,'' meaning that their threshold variables are changed to new
  randomly chosen values.
\item Repeat from step 2.
\end{enumerate}

\begin{figure}
\columnfigure{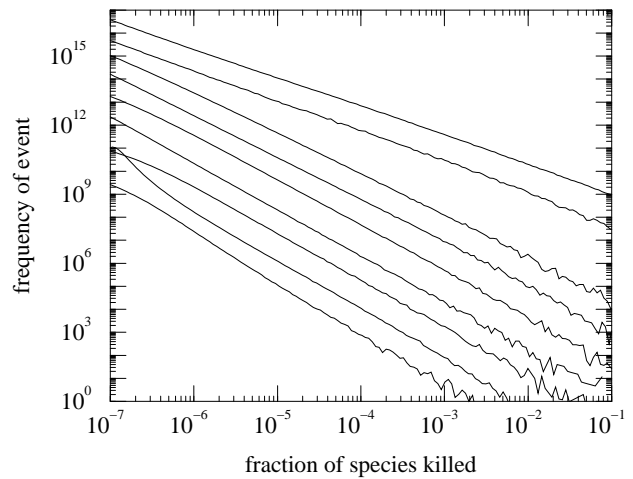}
\caption{The distribution of the sizes of extinction events in
  Newman's model\protect\cite{Newman97} for a variety of different types of
  applied stress including Gaussian centered around zero, Gaussian centered
  away from zero, Poissonian, exponential, stretched exponential, and
  Lorentzian.}
\label{mejnaval}
\end{figure}

The only remaining parameters to be fixed are the value of $f$ and the
distribution $p_{\rm stress}$.  In fact, it turns out that the principal
predictions of the model do not depend on these choices, within reason.
The value of $f$ should be small.  Typical values are on the order of
$10^{-4}$ or less.  The model equilibrates slower for smaller values, but
the results produced are cleaner.  The effect of different choices for
$p_{\rm stress}$ is illustrated in Figure~\ref{mejnaval} where we show the
distribution of the sizes of extinction events in the model---the number of
species that become extinct per time step---for a variety of common noise
distributions, including Gaussian noise, Poissonian noise, and power laws.
As the figure shows, the distribution of event sizes closely follows a
power law, with an exponent of about $-2$ for all of these distributions.
Sneppen and Newman\cite{SN97} have explained this result using an
approximate mean-field-like treatment of the model.  It is possible to
choose a distribution of the applied stresses that will not produce a
power-law extinction size distribution (a uniform distribution between zero
and one will not, for example), but the cases shown in
Figure~\ref{mejnaval} cover most of the distributions likely to be found in
nature.

Newman's model fits in better with the conventional wisdom within the
paleontology community about the causes of extinction events than does the
Bak--Sneppen model, and also produces an exponent for the extinction size
distribution which is close to that observed in the fossil record.
However, it too has its shortcomings.  One of these, which we address next,
is that, like the Bak--Sneppen model, Newman's model is a model of an
equilibrium world in which the average behavior of the ecosystem does not
change over time.  This, as we now discuss, is not the case in real life.

\begin{figure}
\columnfigure{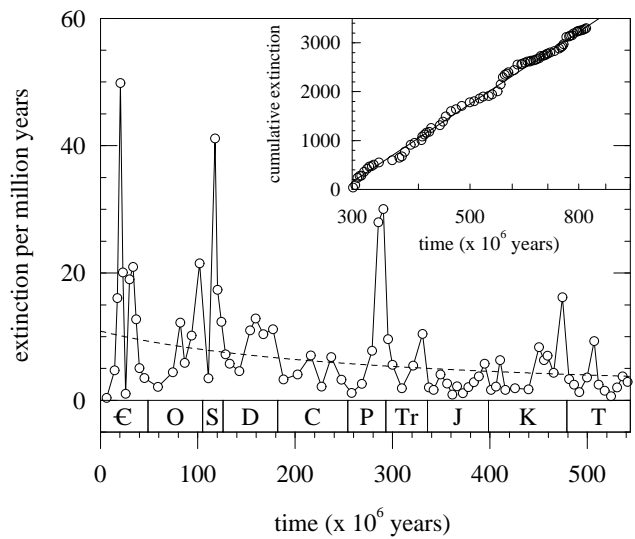}
\caption{Extinction rate as a function of time during the
  last 540 million years.  The dotted line is Eq.~\eref{eqdecline}.  Inset:
  the cumulative extinction, which appears approximately as a straight line
  on the linear-log scale used here.  The data are taken from the
  compilation by Sepkoski.\protect\cite{Sepkoski93}}
\label{decline}
\end{figure}

\section{The model of Sibani, Schmidt, and Alstrom}
In Figure~\ref{decline} we show the rate of extinction of families measured
in each stratigraphic stage as a function of time from 540 million years
ago to the present.  As we can see, the average extinction rate appears to
decline towards the present.  There are significant fluctuations about this
trend---a number of mass extinctions are visible, for instance, as large
peaks in graph---but overall there is a clear decline.  This trend is
believed to be a real effect---species are living longer and becoming
extinct more slowly now than they were a few hundred million years ago
(ignoring recent anthropogenic extinctions).  In the inset to
Figure~\ref{decline}, we show a plot of the {\it cumulative extinction,}
that is, the total number of families (in this case) that disappear from
the data set between its start and a given time $t$ as a function of $t$.
The plot has a logarithmic time axis and when plotted in this way the data
fall on a very nice straight line.  This plot implies that the cumulative
extinction takes the form\cite{NE99}
\begin{equation}
c(t) = A + B\log(t-t_0),
\end{equation}
and the extinction rate $r(t)$, which is the derivative of $c(t)$,
satisfies
\begin{equation}
r(t) = {B\over t-t_0}.
\label{eqdecline}
\end{equation}
Thus the average extinction rate is clearly not constant in time, as the
models of Bak and Sneppen and of Newman implicitly assume.  In fact, it
declines quite sharply.

What implications does this behavior have for the distributions of
quantities such as the sizes of extinction events?  The interval of time
$\Delta t$ in which $r(t)$ falls between $r$ and $r+\Delta r$ is given by
\begin{equation}
\Delta t = {\d t\over\d r} \Delta r,
\end{equation}
and the number of stages or other intervals of time in which the extinction
rate lies in a certain range is proportional to this same expression, that
is, proportional to the derivative
\begin{equation}
{\d t\over\d r} = - {B\over(t-t_0)^2} = - {B\over r^2}.
\end{equation}
In other words, if the extinction rate satisfies Eq.~\eref{eqdecline}, then
the distribution of extinctions in short time intervals such as stages
follows precisely the power law with exponent $-2$ suggested for the fossil
record.  This explanation of the power law is not perfect, because it
assumes that extinction takes precisely the form~\eref{eqdecline}, when in
fact this form is only an average.  More importantly, it really only passes
the buck.  It explains one power law (the distribution of the sizes of
extinction events) by assuming another (the decline in extinction rate).
What is the explanation for this second power law?  A simple model
explaining this behavior has been proposed by Sibani, Schmidt, and
Alstr\o{}m.\cite{SSA95,SSA98}

The model of Sibani~\etal\ is, like the Bak--Sneppen model, based on the
idea of evolution on a fitness landscape.  Again we consider species to be
populations of organisms localized around peaks on the landscape.  And as
before, these populations are considered to be, by and large, stable.  They
change only when a mutation or sequence of mutations takes place which is
large enough to take them to the basin of attraction of a new peak.  In
this model there is no coevolution---the species are considered to be
non-interacting as in Newman's model---but there is one subtlety which is
included that is not present in the Bak--Sneppen model.  If a single
individual in a population has a mutant genotype that puts it in the basin
of attraction of a new peak, then the descendents of that individual may
well evolve toward that new peak.  However, if the fitness at that peak is
lower that the fitness at the peak currently occupied by the rest of the
population, then the mutant population will not supersede the original one,
and no evolution or extinction will take place.  Only if the new peak is
higher than the original one will the population move and the original
species become extinct.

In the model of Sibani~\etal, this process is represented in a very simple
fashion.  Each of $N$ species has a fitness $W_i$.  The process of mutation
to a new peak is represented by generating a random number $r_i$ for each
species $i$ to represent the height of the peak.  If $r_i>W_i$, then the
species evolves and the ancestral species becomes extinct.  Otherwise,
nothing changes.  And that is the entire model.  We can summarize it as
follows:
\begin{enumerate}
\item For each of our $N$ species we choose an initial real fitness value
  $W_i$ at random.  It turns out that it does not matter from what
  distribution we choose these numbers.  The standard thing is to choose
  them uniformly between zero and one.
\item At each time step we choose $N$ new random numbers $r_i$.  All
  species for which $r_i>W_i$ become extinct, and are replaced by
  descendent species which have $W_i=r_i$.
\item Repeat from step 2.
\end{enumerate}
A process of this type is referred to as {\it record
  dynamics}.\cite{Feller60} It is the dynamics one would expect of world
records for any quantity if the values of that quantity fluctuate at random
(which they usually don't).

\begin{figure}
\columnfigure{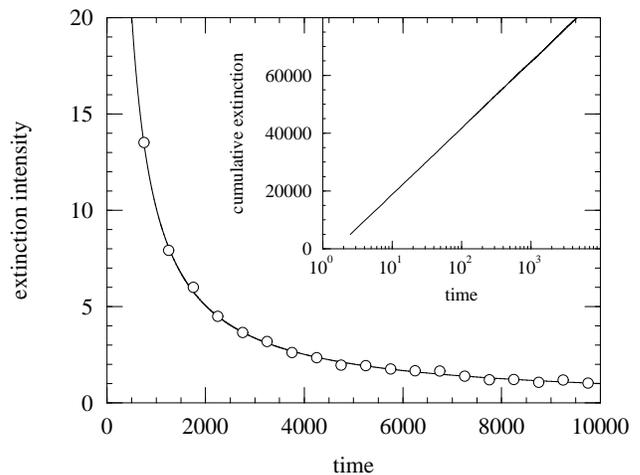}
\caption{The extinction intensity as a function of time for the
  model of Sibani~\etal\protect\cite{SSA95} with $N=10000$.  The points are
  simulation results and the solid line is the expected $1/t$ form.  Inset:
  the cumulative extinction, which appears as a straight line on linear-log
  scales.}
\label{sibani}
\end{figure}

In Figure~\ref{sibani}, we show the results of a simulation of this model
with $N=10000$ species.  The figure has the same layout as
Figure~\ref{decline}: the main figure shows a histogram of the actual
extinction intensity on linear scales, along with the proposed $1/t$ fit;
the inset shows the cumulative extinction.  As we can see, the results
follow the $1/t$ form closely, and the cumulative extinction makes an
excellent straight line on the linear-log scales used, just as in the
fossil record.  It is not difficult to see why this should be the case.
Consider a single species, which after some time $t_0$ has fitness $W_0$.
How long will it take before we generate a random number which is higher
than this value?  On average, it will take the same amount of time that it
took to generate this number the first time, which is $t_0$.  Thus the next
evolutionary event will take place after a total time $t_1=2t_0$.
Repeating the argument, the next one after that will happen at time
$t_2=2t_1=4t_0$, and so on.  In general the $n$th event will happen at
around $t=2^n t_0$.  The number of events $\Delta n$ happening in an
interval of time $\Delta t$ will then be
\begin{equation}
\Delta n = {\Delta t\over t\log 2}.
\end{equation}
In other words, the extinction rate falls as $1/t$.

This model, like the others we have discussed, has its problems.  Chief
among them is the fact that, like the Bak--Sneppen model, it assumes that
all extinction is caused by descendent species superseding their ancestors.
For the case of the large mass extinction events, this is almost certainly
not the true cause of extinction; these events are believed to have been
caused by environmental stress.  However, smaller ``background'' extinction
events do not, by and large, have known causes, so the model of
Sibani~\etal\ is perhaps plausible as a model of background extinction.

\section{Conclusions}
We have outlined three simple models of evolution and extinction which
attempt to explain some of the features seen in the fossil record.  The
model of Bak and Sneppen is a model of extinction caused by large-scale
coevolution---the evolution of one species in response to that of another.
This model is a self-organized critical model that displays ``avalanches''
of coevolutionary activity whose size is distributed according to a power
law.  Newman has proposed a contrasting model in which extinction is caused
by external stresses on the ecosystem.  In this model, species do not
interact at all, but the model still shows a power-law distribution of the
sizes of extinction events.  In the model of Sibani, Schmidt, and
Alstr\o{}m, species evolve when they generate a mutant strain that is
fitter than its parent.  This evolution produces an ever-increasing species
fitness, with jumps, which are associated with extinction events, occurring
less and less frequently over time.  This process also gives rise to a
power-law distribution of extinction events.

So which of these models is right?  Certainly none of them tell the whole
story.  Each one offers a possible explanation of some feature of the
fossil record, but each one leaves out many things as well.  It is is quite
conceivable that all of the mechanisms in these models are occurring
simultaneously in nature and combining to give the signatures we see in the
fossil data.  Or maybe none of them are.  There is a lot of activity in
this field at the moment, and new mechanisms and models are being proposed
all the time.  Models based on ecological interactions, on the structure of
food webs, on competition between species for resources, and on many other
principles are currently under investigation.  Ref.~\citen{Newman99} gives
an extensive review of recent work.  In the long run, it is hoped that
further simulations, along with detailed analyses of the fossil data, will
help us to discover the processes that were at work during the evolution of
life on the Earth.

\section*{Acknowledgements}
The author is grateful to Jack Sepkoski and Doug Erwin for providing data
used in Figs.~\ref{extinction} and~\ref{decline}, and to Per Bak, Gunther
Eble, Doug Erwin, Stuart Kauffman, Paolo Sibani, and Kim Sneppen for useful
discussions.  Harvey Gould and Jan Tobochnik did a nice job with the
editing of the manuscript and made many helpful suggestions.  The work was
supported in part by the Santa Fe Institute and DARPA under grant number
ONR N00014--95--1--0975.

\end{document}